\documentstyle[epsfig,12pt]{article}

\begin{document}

\newcommand{\cc}{\eta_c}
\newcommand{\hef}{ {}^4 He }
\newcommand{\het}{ {}^3 He }
\newcommand{\ra}{\rangle}
\newcommand{\la}{\langle}
\newcommand{\ve}{\varepsilon_0}
\newcommand{\cH}{{\cal{H}}}
\newcommand{\cG}{{\cal{G}}}
\newcommand{\rap}{|\Psi\ra}
\newcommand{\rapp}{|\Psi_0\ra}
\newcommand{\rapt}{|\tilde{\Psi}\ra}
\newcommand{\be}{\begin{equation}}
\newcommand{\ee}{\end{equation}}
\newcommand{\bea}{\begin{eqnarray}}
\newcommand{\eea}{\end{eqnarray}}
\newcommand{\edi}{\mathrm e}

{\large\bf Binding of $\cc$ meson with light nucleus.}\\ [4mm]
{V.B. Belyaev $^a$, A.F. Oskin$^a$, and W. Sandhas$^b$}\\
{\small \em $^a$ BLTP JINR, Dubna, Russia,
$^b$ Physikalisches Institut, Bonn University, Germany}

\date{\today}
\begin{abstract}
Binding energy of $\cc-\hef$ system has been calculated on the basis of Yukawa type $\cc-N$ potential, suggested in [1]. Final Rank Approximation (FRA) has been used for microscopical treatment of the initial 5-body problem. The results are compared with the results of the folding model and variational calculations.
\end{abstract}

\vspace{6mm}

As pointed out in [1] there is an interesting peculiarity in the $\cc-N$
interaction. It turned
 out that the main contribution to the meson-nucleon coupling is due to
multi-gluon exchanges,
a property modelled by a simple Yukawa potential. It means that via the
study of nonrelativistic
systems like $\cc$ plus few nucleons one can get information about the role
of the gluonic
degrees of freedom.

Previous considerations [2] of $\cc$-meson interaction with light nuclei (A =
3,4) were based on
the so called folding model. In this paper, starting from four- and five-
body meson-nuclear
Hamiltonians, we will develop a more elaborate scheme which contains the
folding model as a particular case.

Let us write the total Hamiltonian for the system of one meson and A
nucleons in the form
\be
H = h_0 + \sum_{i = 1}^{A} V_i + H_A \equiv h_0 + V + H_A. \label{totham}
\ee
Here $h_0$ denotes the kinetic energy of the meson relative to the center
of mass of the nuclear
subsystem, $V_i$ is the interaction of the meson with the i-th nucleon, and
$H_A$ the Hamiltonian
of the nuclear subsystem.

In what follows we are interested only in calculating the binding energy
of meson with $\hef$. Having in mind, that in such a situation we are comparatively
far away from the nearest
break-up thresholds for nuclear subsystem, we can approximate
in eq.(\ref{totham}) the nuclear subsystem Hamiltonian $H_A$ by a finite
rank operator, for instance, by the rank-one term
\be
H_A \approx \tilde{H_A} = \ve |\chi_0\ra \la\chi_0|, \label{fra}
\ee
where $\ve$ and $\chi_0$ denote the ground state energy and the
corresponding wave function of
$\het$ or $\hef$ respectively. This Finite Rank Approximation (FRA) is more
general, than the
Resonating Group Method (RGM), as will be seen below from the expression for the
eigenfunctions of
the Hamiltonian (\ref{totham}).

Using eq.(\ref{fra}) one can easily obtains the following represantation
for the approximate solution $\rap$
of the initial Schr\"odinger equation
\bea
\tilde{H} \rapt = E \rapt, \\
\tilde{H} = h_0 + V + \tilde{H}_A \\
\rapt = \ve \cG(E)|\chi_0\ra |B\ra. \label{Psirep}
\eea
Here $\cG(E) = (E - \cH)^{-1}$, $\cH = h_0 + \sum_{i = 1}^{A} V_i$. The
state $| B \ra \equiv \la \chi_0 \rapt$, more explicitly written as $\la \vec{R}\la \chi_0 \rapt$
is a new unknown
function, which by definition depends only on the relative distance
$\vec{R}$ between the meson and the center of mass of the nucleonic subsystem.

From the structure of the hamiltonian $\cH$ it is clear, that the Green
function $G(E)$ depends
on $\vec{R}$, as a dynamical variable, and parametrically on the
internuclear distances. This observation means, that expression (\ref{Psirep}) for the total solution
never can be presented by the widely used RGM ansatz
\be
|\Psi_{RG}\ra = |\chi_0\ra |B\ra, \label{partform}
\ee
Only with additional approximation of type
\be
\la\chi_0| \cG(V) | \chi_0\ra \approx \la\chi_0 | \cG(\overline{V}) | \chi_0 \ra,
\ee
where $\overline{V} \equiv \la\chi_0|V|\chi_0\ra$, one can reduce the general presentation (\ref{Psirep}) to the particular RGM form (\ref{partform}).

Projecting eq.(\ref{Psirep}) onto $| \chi_0\ra$ one can immediately come to the one-dimensional integral equation for the unknown  function $|B\ra$. However, the kernel of this sort of equation can be calculated analytically only for the separable type of meson-nucleon potentials $V_i$. Since we work with local potentials of Yukawa type, it is more practical to apply a different procedure. For this purpose, let us introduce the set of $|\chi_i \ra$ \footnote{For details see the Appendix} auxiliary orthonormal square integrable functions defined in nuclear subspace. As the first term of this set let us take the wave function of the ground state of nuclei $\hef$, $| \chi_0 \ra$
\be
	H_A |\chi_0\ra = \ve |\chi_0\ra.
\ee
Inverting Green function in eq.(\ref{Psirep}) and using the expansion
\be
	|\Psi\ra = \sum_{i = 0}^N |B_i\ra \cdot |\chi_i\ra,
\ee
one can obtain the system of one-dimensional differential equations for the functions $|B_i\ra$
\be
	\sum_{i = 0}^N \la \chi_k|\cH - E |\chi_i\ra |B_i\ra = -\ve|B_k\ra \delta_{k0}. \label{difsys}
\ee
It is clear that if one restricts oneself only to the first equation of the system (\ref{difsys}), the folding model for interaction of meson with nucleus will appear.

Now let us present the results of calculation of binding energy of the $\cc$-meson with the nucleus $\hef$. As it was suggested in [1], meson-nucleon interaction can be described by the attractive Yukawa potential
\be
	V_{mN}(r) = -V_0 \frac{\edi^{-\mu r}}{r},
\ee
with parameters varying in the following intervals:
\bea
	&&0.4 \le V_0 \le 0.6 \nonumber \\
	&&0.4 \le \mu \le 0.6 (GeV)
\eea
In accordance with the results of [2], the lowest approximation, corresponding to the folding model (solid line on Fig.[1]), produces the bound state in the system. It is clear from the Fig.[1] that there is rather a fast convergence of the suggested procedure for all values for strengths of the meson-nucleon potential. In other words, it means that Green function
\be
	\cG (E) = (E - \cH)^{-1},
\ee
which describes a relative motion of meson and the center of mass of the nucleus, can be reasonably well represented for the eigenvalue problem by a finite matrix of low dimension.

The comparision of numerical values of the binding energy in the present approach and in the folding model [2] gives $E_{\rm fold}^{({{\rm a}=0.6, \mu = 0.6 {\rm GeV}})} = - 5 {\rm MeV}$, $E_{\rm present}^{({{\rm a}=0.6, \mu = 0.6 {\rm GeV}})} = - 7.5 {\rm MeV}$. Underestimation of binding in the folding model is  due to neglecting in this model of rescattering of meson by nucleons of the target. In our formulation, this effectes fully taken into account in the Green function $\cG(E)$.

In [1] the binding energy we are interested in appeared to be equal to $E_{\rm phen}~=~-~143{\rm MeV}$ which is probably related with a not fully adequate A-dependence of the meson-nucleus potential, postulated by the authors.

\vspace{1cm}
{\large \bf Appendix }
\vspace{1cm}

We will build the set of auxiliary orthonormal functions $|\chi_i \ra$ in a usual manner. First of all we will take a set of gaussian functions $|\tilde \chi_i\ra$
\be
	|\tilde \chi_i (\alpha_i)\ra = \prod_{j = 1}^{N-1} (\frac{\alpha_{ij}}{\pi})^{3/4} \edi^{-\frac{1}{2} \alpha_{ij} r_j^2},
\ee
where N is a number of the particles in the system, $r_i$ is a jacobian coordinates in the nuclear center of mass frame. As it is obvious this functions have a unit norm, $\alpha_i$ is a set of coefficients $\{\alpha_{i1}...\alpha_{i N - 1}\}$. This coefficients are chosen in the following manner, $\alpha_0$ is chosen so, that $\la \tilde{\chi}_0| r^2 | \tilde{\chi}_0 \ra = r^2_{msr}$, where $r_{msr}$ is a mean square radius of the corresponding nuclea A. For example, if $N = 4$, then this coefficients would be 
\bea
	\alpha_{01} = \frac{1}{2 \la r_{msr}^2 \ra} \nonumber \\
	\alpha_{02} = \frac{1}{2 \la r_{msr}^2 \ra} \nonumber \\
	\alpha_{03} = \frac{1}{\la r_{msr}^2 \ra} \nonumber.
\eea
Coefficients $\alpha_i$ are chosen recursively $\alpha_{i+1} = \alpha_i * C$, where C is some numerical coeffiecient chosen in such a way, that each following function $|\chi_{i + 1}\ra$ would be wider than the previous one. 

Now functions $|\chi_i\ra$ are obtained in a Gramme-Schmidt recursive procedure
\bea
	&&|\chi_1\ra = |\tilde \chi_1\ra \nonumber\\
	&&|\chi_i\ra = \frac{|\tilde\chi_i\ra - \sum_{j = 1}^{i - 1}\la\tilde\chi_i|\chi_j\ra|\chi_j\ra}{(1 - \sum_{j = 1}^{i - 1}\la \tilde\chi_i|\chi_j\ra^2)^{1/2}}.
\eea
Yukawa potential between meson and the j-th particle of the target is taken in the form
\bea
	V(\rho_j)=V_0\frac{\edi^{-\mu\rho_j}}{\rho_j} \nonumber\\
	\rho_j = |\vec{\rho} + \sum_{i = 1}^N \lambda_{ij}\vec{r_i}|,
\eea
where $\vec{\rho}$ is a distance between meson and nuclei center of mass, coefficients $\lambda_{ij}$ specifies position of the corresponding j-th particle. One could see that the following matrix elements takes form
\bea
&&\la\tilde\chi_i(\alpha_i)|V(\rho_k)|\tilde\chi_j(\alpha_j)\ra = \frac{V_0}{2 \sqrt{\pi}}\int_0^{\infty}dt\frac{\edi^{-\frac{\rho^2}{4(t + \gamma_{kij})}-t\mu^2}}{(t + \gamma_{kij})} \nonumber\\
&&\gamma_{kij} = \sum_n\frac{\lambda_{nk}^2}{2(\alpha_{in} + \alpha_{jn})}.
\eea
To obtain formulas for $\la\chi_i|V|\chi_j\ra$ it's very convinient to use following recursive formulas
\bea
	 \la\chi_i|V|\chi_j\ra =  \la\chi_j|V|\chi_i\ra = \frac{1}{(1 - \sum_{k = 1}^{j - 1}\la\tilde\chi_j|\chi_k\ra^2)^{1/2}}( \la\chi_i|V|\tilde\chi_j\ra - \sum_{k=1}^{j - 1} \la\chi_k|\tilde\chi_j\ra \la\chi_i|V|\chi_k\ra)\nonumber\\
	 \la\chi_i|V|\tilde\chi_j\ra = \frac{1}{(1 - \sum_{k = 1}^{i - 1}\la\chi_k|\tilde\chi_i\ra^2)^{1/2}}( \la\tilde\chi_i|V|\tilde\chi_j\ra - \sum_{k=1}^{i - 1}  \la\chi_k|\tilde\chi_i\ra \la\tilde\chi_j|V|\chi_k\ra).
\eea
Due to the first equality we could always assume that $i \le j$.

Results for the binding energy of $\cc \hef$ system are presented on the figure below.
\begin{figure}[h]
    \includegraphics[width=14cm,clip,keepaspectratio]{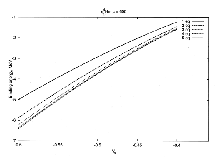} \label{fig1}
\end{figure}


\begin{thebibliography}{99}

\bibitem{Brodsky} Brodsky S.J. et al., PRL {\bf 64}, 1011 (1990)
\bibitem{Wasson} Wasson D.A. PRL {\bf 67}, 2237 (1991)

\end{thebibliography}
\end{document}